\documentclass[12pt]{iopart}

\newcommand\eg{e.g.\ }

\usepackage{bm}
\usepackage{graphicx}
\usepackage{harvard}
\usepackage{url}

\begin{document}

\title[]{Leading-edge vortex shedding from rotating wings}

\author{Dmitry Kolomenskiy$^1$\footnote{Corresponding
author: dkolom@gmail.com}, Yossef Elimelech$^2$ and Kai Schneider$^3$}

\address{$^1$Centre de Recherches Math\'ematiques (CRM),
Department of Mathematics and Statistics, McGill University, 805 Sherbrooke W., Montreal, QC, H3A 0B9, Canada}
\address{$^2$Faculty of Aerospace Engineering, Technion-Israel Institute of Technology, Haifa 32000, Israel}
\address{$^3$M2P2--CNRS, Universit\'e d'Aix-Marseille, 39, rue Fr\'ed\'eric Joliot-Curie, 13453 Marseille Cedex 13, France}

\ead{dkolom@gmail.com}

\begin{abstract}
The paper presents a numerical investigation of
the leading-edge vortices generated by rotating triangular wings at Reynolds number $Re=250$.
A series of three-dimensional numerical simulations have been carried out using a Fourier pseudo-spectral method with volume penalization.
The transition from
stable attachment of the leading-edge vortex to periodic
vortex shedding is explored, as a function of the wing aspect
ratio and the angle of attack.
It is found that, in a stable configuration, the spanwise flow in the recirculation bubble past the wing
is due to the centrifugal force, incompressibility and viscous stresses.
For the flow outside of the bubble, an inviscid model of spanwise flow is presented.
\end{abstract}

\vspace{2pc}

\maketitle

\section{Introduction}\label{Sec:intro}

Vortices and vorticity production play an important role in
insect flight. 
Insect wings have rather sharp leading edges at which the flow
separates, thus producing a large amount of vorticity. The
unsteadiness and the three-dimensional character of the flow lead
to complex vortex dynamics and interactions with the wings. It is
known that some of them have a strong positive effect on the
generation of lift \citeaffixed{Lehmann_2004}{e.g.}. Notably, the
three-dimensional character of the flow changes the dynamics of
the vortex shedding from flapping or revolving wings, compared to
the case of rectilinear motion, in the range of Reynolds
number typical of insect flight \cite{Liu_Kawachi_2001}. Even
when the wings operate at large angles of attack, the vorticity
generated at the leading edge remains above the suction side (upper surface) of the wing. The pressure deficit in the recirculation
bubble results in high lift at large angles of attack. This feature makes a striking
contrast to the periodic vortex shedding that occurs in a
two-dimensional motion. This `stable' behaviour of the
leading-edge vorticity is
accompanied by a strong flow in the spanwise direction from the wing root towards its tip \cite{Maxworthy_79}. 

To explore these effects, we have carried out three-dimensional Navier--Stokes simulations
using a pseudo-spectral method with volume penalization
\cite{Kolomenskiy_etal_11b}. These simulations are validated
against experimental analysis in our earlier publication \cite{Elimelech_etal_2013}. 
We consider a wing revolving about the vertical axis. The wing has
a cross-section of a flat plate and its planform is triangular
such that the ratio of the local radius $r$ to the chord length
$c(r)$ is constant along the wing span (see
figure~\ref{fig:schematic}). \citeasnoun{Lentink_Dickinson_09} suggest
that the stability of the leading-edge vortices depends on the ratio $r/c$.
Indeed, high aspect ratio wind turbine blades usually stall first
along the distal portion of the blade, while near the hub the flow
remains attached and highly three-dimensional (having a large
spanwise velocity component). \citeasnoun{Lentink_Dickinson_09} also associate
the very large local sectional lift coefficient to the
fact that the flow is highly three-dimensional in the hub region.
The lift coefficient at the root
sections can reach very large values. Meanwhile, experiments with
revolving models of insect wings by \citeasnoun{Usherwood_Ellington_02a}
and recent numerical simulations by \citeasnoun{Harbig_etal_2012} did
not reveal any strong influence of the aspect ratio, probably
because the shape of those wings was such that locally $r/c$ was
small enough at the sections near $r/R=0.5$...0.7, which produce most
of the net lift force. Only the wing tip region showed sensitivity
to the aspect ratio.
\citeasnoun{Harbig_etal_2013} also suggested
that the flow structure is determined by the span-based Reynolds number
rather than the chord-based Reynolds number. However, this
scaling only holds if the spanwise velocity component is large above the wing surface.

The triangular wing shape considered in this
work has the advantage that the flow in all cross-sections is
characterized by the same value of the local aspect ratio
$r/c=R/C$, where $R$ is the wing length and $C$ is the tip chord.
Therefore, cross-sections in the middle portion of the wing
operate under similar conditions.
Our results indicate that very elongated wings do not generate stable leading-edge vortices, unlike less elongated wings.
We identify the marginal value of the local aspect ratio corresponding to this transition.

Moreover, it should be noted that the properties of the leading-edge vortex
significantly vary as the Reynolds number varies \cite{Shyy_Liu_2007}.
Our work focuses on wings operating at the Reynolds number $Re=250$, representative of smaller insects,
e.g. it is about 130 for a fruit fly and about 480 for a mosquito.

In section~\ref{Sec:numerical_setup} we describe the flow configuration and briefly recall the numerical method.
Numerical results 
and a potential flow model for the spanwise flow are discussed in section~\ref{Sec:results}.
Finally, conclusions and perspectives are presented in section~\ref{Sec:conclusion}.

\section{Numerical setup}\label{Sec:numerical_setup}

\begin{figure}
\centering
\includegraphics[scale=1.0]{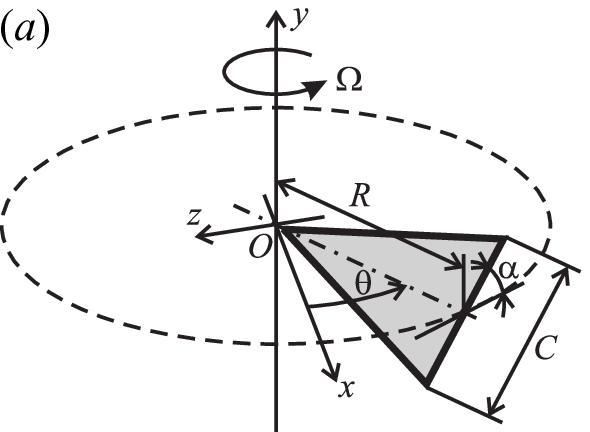} \quad
\includegraphics[scale=1.0]{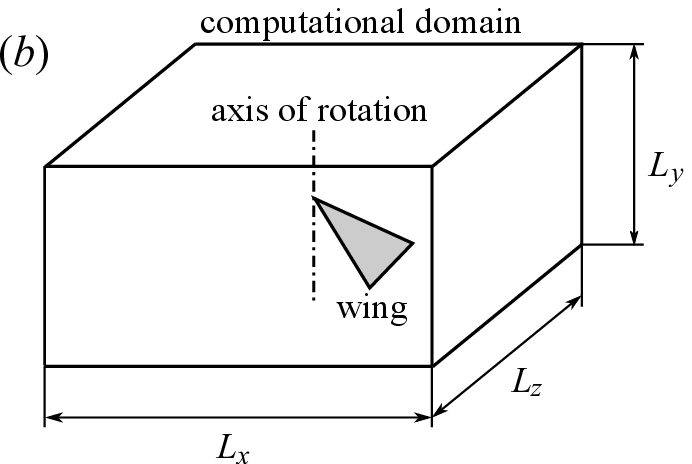}
\caption{ (\textit{a}) Schematic diagram showing the wing rotating about the
vertical axis. (\textit{b}) Computational domain.} \label{fig:schematic}
\end{figure}

We consider a single wing rotating about vertical axis $Oy$, as
shown in figure~\ref{fig:schematic}. The wing is a flat plate of
thickness $h=0.12c_{0.5}$, where $c_{0.5}$ is the mid-span chord length.
It is inclined with respect to the horizontal plane
$Oxz$ at angle $\alpha$ that we call the angle of attack. The wing
shape is triangular, such that the chord length $c$ varies
linearly with radius $r$: $c(r)=2r/\Lambda$, where $\Lambda$ is
the aspect ratio conventionally defined as the square of the wing
length divided by the the area \citeaffixed{Anderson_2000}{see, e.g.}. In this particular
case, we have $\Lambda/2=R/C$. 
Moreover, the local ratio $r/c(r)$ is constant and equal to
$\Lambda/2$.

The motion starts from rest at $t=0$ and the angular velocity
$\dot{\theta}$ varies like
\begin{equation}\label{kinematics}
\dot{\theta}/\Omega = 1-e^{-t/\tau}.
\end{equation}
Equation (\ref{kinematics}) results in a gradual increase of the angular velocity
until it reaches 99\% of its ultimate value at $t/\tau = -\ln{0.01}$, then it remains approximately constant.
The gradual increase of the velocity allows avoiding singularity of the aerodynamic force at $t=0$, otherwise present if the motion starts impulsively.
In order to obtain comparable
vortex shedding frequencies in computations with different values
of $\Lambda$, it is convenient to ensure that, in all cases, the
mid-span section travels the same distance per unit time. This
implies that the time evolution of $\theta/(2\arctan{\Lambda/2})$
should be the same in all cases. We satisfy that approximately
by setting $\Omega = 4/\Lambda$.
The scaling factor $\tau=0.4348$ is expressed in the
dimensionless units explained hereunder.
Thus, the angular velocity reaches 99\% of its ultimate value at time $t=2$, which is the same for all cases.
The corresponding angular position varies between cases.
In the case $\Lambda=4$, it corresponds to $\theta=90^\circ$, consistently with our earlier study
\cite{Elimelech_etal_2013}.
In this work, we are not interested in the initial transient and our analysis of the flow only focuses on time $t>2$,
when the angular velocity $\dot{\theta}$ is approximately constant in time.

The computational code operates dimensionless quantities, and it
is also convenient to present the results using the same
normalization. The air density is constant and equal to unity,
$\rho=1$. All distances are normalized to the mid-span chord
length, i.e., $c(R/2)=c_{0.5}=1$. Velocities are normalized to $\Omega R/4$.
The Reynolds number is based on these two quantities and the
kinematic viscosity $\nu$,
\begin{equation}\label{reynolds number}
Re = \frac{\Omega R/2 \cdot c_{0.5}}{\nu} = \frac{2}{\nu},
\end{equation}
so that it is independent of $\Lambda$.

Our Navier--Stokes solver is described in earlier publications \cite{Kolomenskiy_etal_11b,Kolomenskiy_Schneider_09} in a greater detail.
The incompressible three-dimensional Navier--Stokes equations are solved using a Fourier pseudo-spectral method.
The no-slip boundary condition at the wing is imposed using the volume penalization method \cite{Angot_etal_1999}.
Our approach to modelling moving obstacles is described in \cite{Kolomenskiy_Schneider_09}.
Our parallel implementation of the code is based on FFTW \cite{FFTW05} and P3DFFT \cite{Donzis_etal_2008} fast Fourier transform packages.

\section{Results and discussion}\label{Sec:results}

\subsection{Vortex dynamics and aerodynamic forces}\label{Sec:dynamics_forces}

Let us first compare three numerical simulations with $\Lambda=4$, 8 and 16.
The domain sizes are, respectively, $L_x \times L_z \times L_y = 12.75^3$, $21^2 \times 10.5$ and $34^2 \times 10.625$.
The discretization grids are uniform Cartesian.
They contain $384^3$, $640^2 \times 320$ and $1024^2 \times 320$ points, respectively.
Therefore, the grid step size is approximately the same in all cases.
All wings operate at $\alpha=30^\circ$ and $Re=250$.

\begin{figure}
\centering
\includegraphics[scale=0.80]{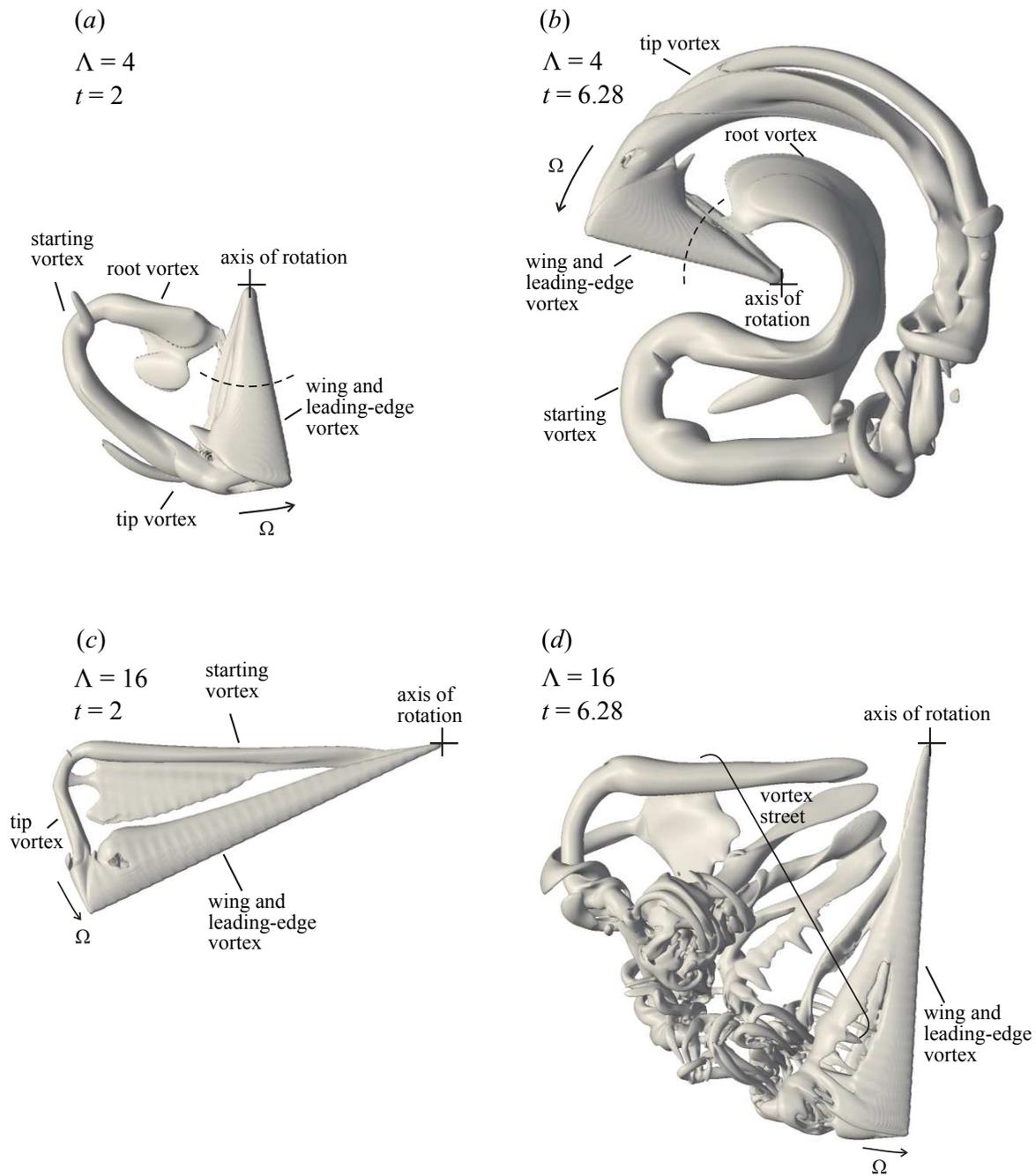}
\caption{Isosurfaces of the $\lambda_2$-criterion 
for low and high aspect ratio wings: \rm{(}\it{a,b}\rm{)} $\Lambda=4$ and \rm{(}\it{c,d}\rm{)}
$\Lambda=16$. Two different time instants are shown: \rm{(}\it{a,c}\rm{)} $t=2$ and \rm{(}\it{b,d}\rm{)} $t=6.28$. Corresponding position angles are equal to \rm{(}\it{a}\rm{)} $\theta=90^\circ$, \rm{(}\it{b}\rm{)} $335^\circ$, \rm{(}\it{c}\rm{)} $22.5^\circ$ and \rm{(}\it{d}\rm{)} $83.8^\circ$. Figures for $\Lambda=4$ and $\Lambda=16$ are not at the same scale.}\label{fig:lam2}
\end{figure}

It is instructive to consider the vortex system of a rotating wing
before analyzing the dynamics of the leading-edge vorticity. As an
indicator of vortex cores, it is convenient to use the
$\lambda_2$-criterion introduced by \citeasnoun{Jeong_Hussain_95}.
The vortex is defined as the region of $\lambda_2<0$,
 $\lambda_2$ being the second largest eigenvalue of $\bm{S}^2+\bm{\Omega}^2$,
where $\bm{S}$ and $\bm{\Omega}$ are, respectively,
the symmetric and antisymmetric components of the velocity gradient tensor $\nabla\bm{u}$.
Note that the zero isosurface does not belong to the vortex.
Iso-surfaces of $\lambda_2=-0.01$ (an arbitrary negative value close to zero) are shown in figure~\ref{fig:lam2} for two different
flows: past a low aspect-ratio wing ($\Lambda=4$) and a high
aspect-ratio wing ($\Lambda=16$).

Panels \rm{(}\it{a}\rm{)} and \rm{(}\it{b}\rm{)} show the wake of the low aspect ratio wing, $\Lambda=4$, at
two different time instants, $t=2$ and $6.28$, respectively.
Time $t=2$ corresponds to the end of the initial acceleration phase.
The wake has a typical structure of the finite aspect ratio: a starting vortex, a tip vortex and a root vortex.
The leading edge vortex
is situated above the wing and has approximately conical shape,
i.e., all wing sections operate under similar conditions.
At $t=6.28$, the leading edge vortex remains attached and travels with the wing.
The wake preserves its original closed-loop structure, despite some instability near the wing tip.

Panels \rm{(}\it{c}\rm{)} and \rm{(}\it{d}\rm{)} show the wake of the high aspect ratio wing, $\Lambda=16$, at the same time instants.
At $t=2$, the same closed-loop structure is visible as in the previous case, except that the root vorticity is now more diffuse, hence not visualized.
However, by $t=6.28$,
the wing generates an array of radial vortices that
separate from the leading and trailing edges.
Those are not present in the wake of the low aspect-ratio wing.
This fundamental difference is the main focus of our work.

\begin{figure}[p]
\centering
\includegraphics[scale=1.1]{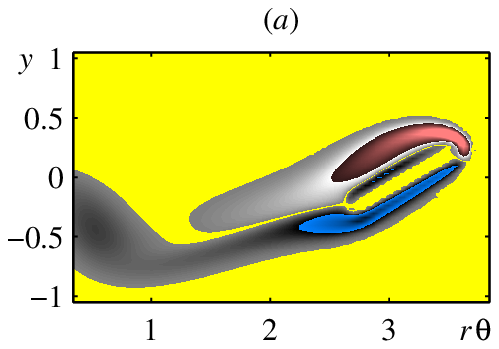}\hspace{0.5cm}
\includegraphics[scale=1.1]{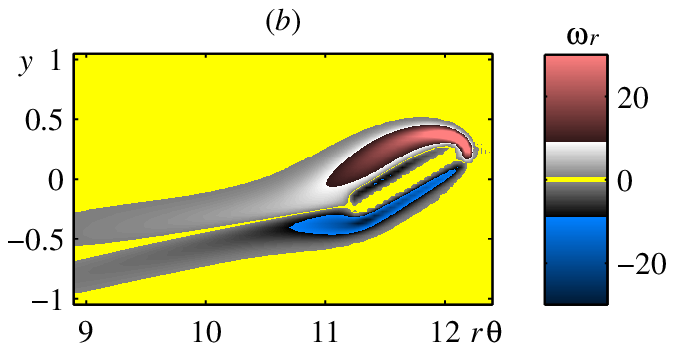}
\caption{Radial component of the vorticity in the mid-span section
($r/R=0.5$) for $\Lambda=4$. Time instants are \rm{(}\it{a}\rm{)}
$t=2$ and \rm{(}\it{b}\rm{)} $t=6.28$. Position angles are
$\theta=90^\circ$ and $335^\circ$, respectively.}\label{fig:vorticity_lam4}
\end{figure}
\begin{figure}[p]
\centering
\includegraphics[scale=1.1]{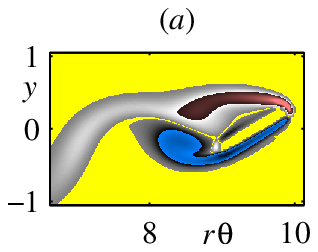}\hspace{0.5cm}
\includegraphics[scale=1.1]{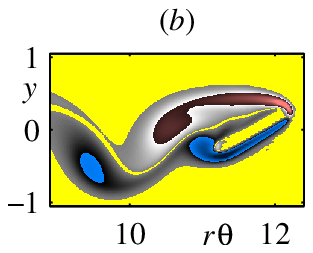}\hspace{0.5cm}
\includegraphics[scale=1.1]{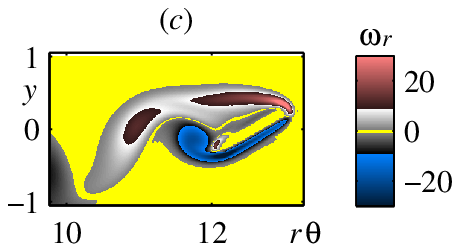}
\caption{Radial component of the vorticity in the mid-span section
($r/R=0.5$) for $\Lambda=16$. Time instants are \rm{(}\it{a}\rm{)}
$t=5.14$, \rm{(}\it{b}\rm{)} $t=6.28$ and \rm{(}\it{c}\rm{)}
$t=6.71$.
Position angles are
$\theta=67.5^\circ$, $83.8^\circ$ and $90^\circ$, respectively.}\label{fig:vorticity_lam16}
\end{figure}
\begin{figure}[p]
\centering
\includegraphics[scale=1.1]{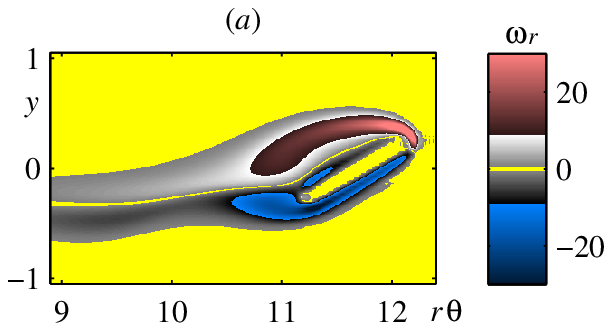}\hspace{0.5cm}
\includegraphics[scale=1.1]{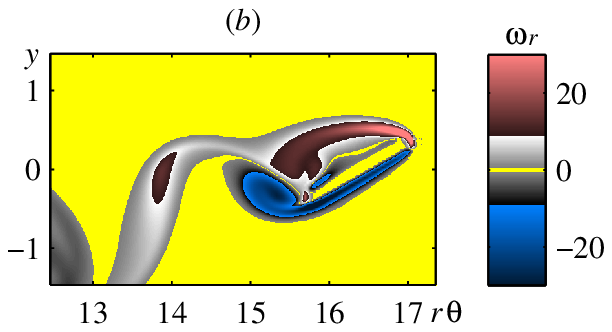}
\caption{Radial component of the vorticity for $\Lambda=8$ in two
sections: \rm{(}\it{a}\rm{)} $r/R=0.5$ and \rm{(}\it{b}\rm{)}
$r/R=0.7$. Time $t=6.28$, position angle $\theta=167.5^\circ$.}\label{fig:vorticity_lam8}
\end{figure}

A more detailed view of the leading-edge vorticity dynamics can be
obtained from two-dimensional sections.
Figure~\ref{fig:vorticity_lam4} shows two snapshots of the radial
component of the vorticity in a cylindrical section of radius
$r/R=0.5$, as indicated with a dashed line in
figure~\ref{fig:lam2}. The aspect ratio is $\Lambda=4$. The flow
separates from both leading and trailing edges, and two layers of
counter-rotating vorticity emerge from the shear layers between the
outer flow and the recirculation bubble past the wing. These
vortices grow as the wing accelerates and approach their
equilibrium state as the angular velocity $\dot{\theta}$
approaches its ultimate value. Thus, at later time, the vorticity
pattern displayed in figure~\ref{fig:vorticity_lam4}(\textit{b})
remains unaltered until the end of the first revolution, then it
is slightly modified by the downwash velocity induced by the tip
vortex. Note that, during hovering, the downwash is present if
the wings flap as well as if they rotate.
For flapping wings during hovering,
periodic time evolution of the aerodynamic force
establishes after 2 or 3 strokes \cite{Wang_etal_2004}.
Interaction of the leading-edge vortex with the wake of the preceding strokes
is complex. It consists not only in the downwash effect,
but also in wing-vortex interactions. This complicates the matter of
the leading-edge vortex stability, since the interactions are strong and intrinsically unsteady.
We follow a reductionist approach and, in this work, only focus on
the development of leading-edge vortices in an undisturbed environment.
Therefore, we only consider the first wing revolution.
It may be conjectured that perturbations due to wing-wake interactions
are an additional destabilizing secondary effect.

\begin{figure}[p]
\centering
\includegraphics[scale=0.67]{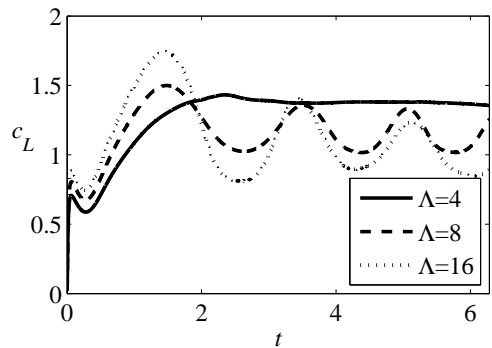}
\caption{Time evolution of the lift coefficient of different wings
at the angle of attack $\alpha=30^\circ$.} \label{fig:lift}
\end{figure}

Mid-span radial vorticity plots in the case $\Lambda=16$
are shown in figure~\ref{fig:vorticity_lam16}. At the
instants shown in the figure, $\dot{\theta}$ is
close to its ultimate value $\Omega$. The figure clearly indicates vortex shedding. As
one can see in figure~\ref{fig:lam2}(\textit{b}), these vortices
extend in the spanwise direction from $r/R=0.25$ to $r/R=1$
shortly after separation, then they are entrained by the tip
vortex and roll up.

The intermediate case $\Lambda=8$ exhibits both scenarios,
depending on the spanwise location. The vortices remain attached
at $r/R<0.6$ and shedding occurs at $r/R>0.6$, as shown in
figure~\ref{fig:vorticity_lam8}.

Figure~\ref{fig:lift} displays the time evolution of the lift
coefficient, defined by $c_L=2L/\rho U^2_{0.65}A_{wing}$, where
$U_{0.65}=0.65 \Omega R$, $A_{wing}=RC/2$. In this notation, $L$
is the dimensional lift force and $\Omega$ is the angular velocity.
The choice of the reference velocity $U_{0.65}$ at $r=0.65R$ follows from the blade
element theory: it ensures that the lift coefficient of the wing is about the
same as the average section lift coefficient.
The lift
coefficient of the $\Lambda=4$ wing is almost constant after an
initial transient. The $\Lambda=16$ wing exhibits a large
overshoot during the initial acceleration, drops down to a half of
the maximum value and continues oscillating with smaller
amplitude. The peak values also become smaller, and the
time-averaged long-term lift coefficient is lower than the one of
the low aspect ratio wing. The curve corresponding to $\Lambda=8$
is situated in-between, because the unsteady vortex shedding
occurs only from the distal portion of the wing.

\begin{figure}
\centering
\includegraphics[scale=1.1]{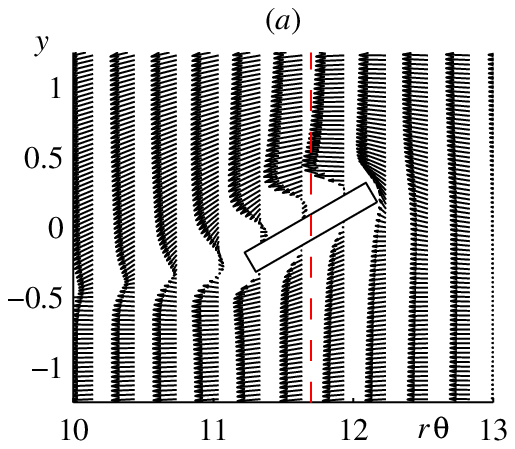}
\includegraphics[scale=1.1]{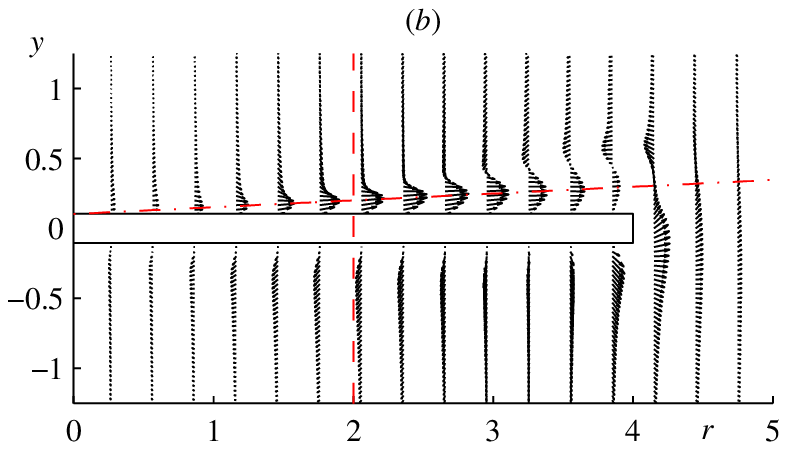}
\includegraphics[scale=1.1]{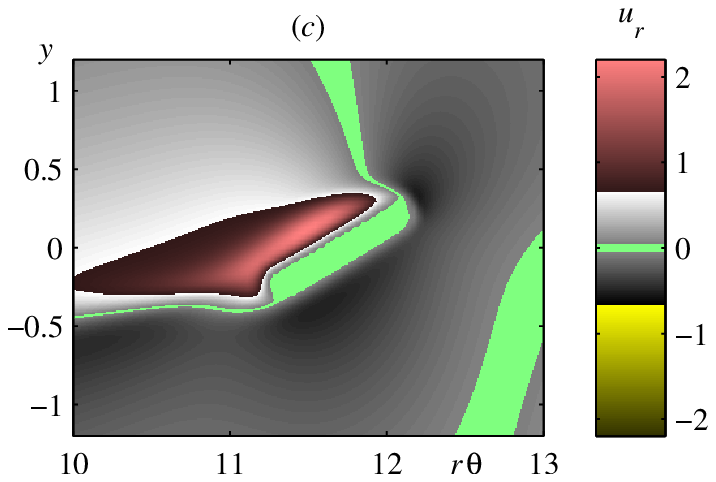}
\includegraphics[scale=1.1]{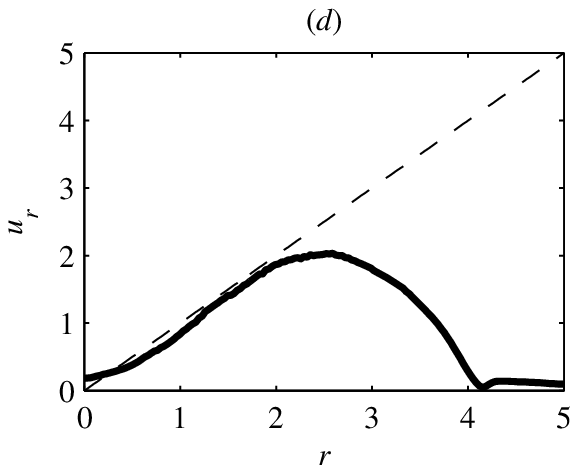}
\caption{Velocity field for $\Lambda=4$. In-plane velocity in \rm{(}\it{a}\rm{)}
mid-span section and \rm{(}\it{b}\rm{)} spanwise section planes. Each
of the two views also shows the position of the other section,
indicated by a red dashed line. Vectors are at the same scale in both
panels. 
\rm{(}\it{c}\rm{)} Radial velocity component in the mid-span section. 
\rm{(}\it{d}\rm{)} Radial velocity component along the horizontal red dash-dotted line indicated in
figure \rm{(}\it{b}\rm{)}. The black dashed line shows the reference circumferential velocity $\Omega r$.
Time $t=6.28$, position angle $\theta=335^\circ$.}\label{fig:velocity}
\end{figure}


The large lift coefficient of the low aspect ratio wing, $\Lambda=4$,
is due to the pressure deficit in the recirculation bubble
\cite{Corten_01}. This recirculation bubble is visible in
figure~\ref{fig:velocity}(\textit{a}) as a mass of fluid above the
upper surface having zero relative velocity in the chordwise
plane. Centrifugal forces drive this fluid in the spanwise
direction towards the wing tip (see
figure~\ref{fig:velocity}\textit{b}).
As shown in figure~\ref{fig:velocity}(\textit{c}),
the maximum spanwise velocity is in the centre of recirculation bubble,
and it is of the same order of magnitude as the local reference velocity $\Omega r$
(see also figure~\ref{fig:velocity}\textit{d} that shows spanwise variation
of the velocity inside the bubble).

\begin{figure}
\centering
\includegraphics[scale=1.1]{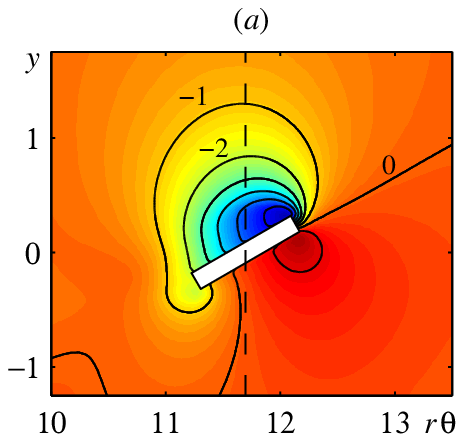}\hspace{0.5cm}
\includegraphics[scale=1.1]{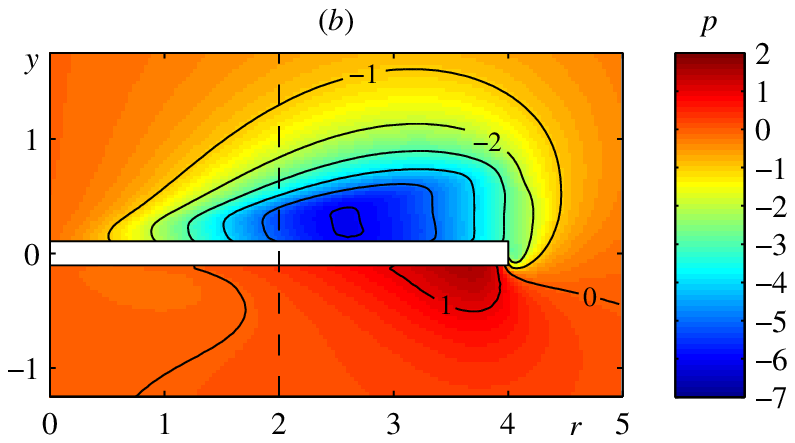}
\caption{Iso-contours of the pressure for $\Lambda=4$, plotted with step 1.
\rm{(}\it{a}\rm{)} mid-span section and \rm{(}\it{b}\rm{)}
spanwise section. Each of the two views also shows the
position of the other section, indicated by a dashed
line. Time $t=6.28$, position angle $\theta=335^\circ$.}\label{fig:pressure}
\end{figure}

Isobars in a chordwise and in a spanwise section are
shown in figure~\ref{fig:pressure}. Unlike in a recirculation
bubble behind a bluff body in a translational motion, here the
static pressure deficit on the upper surface is at least twice as
large as the local (at $r=R/2$) reference dynamic pressure, $0.5 \rho (\Omega
R/2)^2$. The largest negative pressure is in the vortex core near
the leading edge and it gradually decreases in magnitude towards
the wing's trailing edge. These observations are in agreement with Corten's theory \cite{Corten_01}.
In the spanwise direction, the pressure deficit in the
recirculation bubble grows between $r=0$ and $2.6$, then decreases
towards the wing tip (see figure~\ref{fig:pressure}\textit{b}).

Let us consider the radial component of the momentum equation in cylindrical polar coordinates,
in a reference frame rotating with angular velocity $\Omega$ about the vertical axis $y$:
\begin{equation}
\frac{\partial u_r}{\partial t} + \bm{u}\cdot\nabla u_r - \frac{u_\theta^2}{r} + \frac{1}{\rho}\frac{\partial p}{\partial r} - \nu \left( \nabla^2 u_r - \frac{u_r}{r^2} - \frac{2}{r^2} \frac{\partial u_\theta}{\partial \theta}  \right) = 2 \Omega u_\theta + \Omega^2 r.
\label{eq:momentum_polars}
\end{equation}
The flow in the recirculation bubble is approximately steady, therefore $\partial u_r / \partial t \approx 0$.
In the centre of the bubble, velocity components $u_\theta$ and $u_y$ and their first derivatives are small (see figure~\ref{fig:velocity}\textit{a}).
In particular, this implies that the Coriolis acceleration $2 \Omega u_\theta$ is small.
Hence, in the centre of the bubble, equation (\ref{eq:momentum_polars}) can be simplified to
\begin{equation}
u_r \frac{\partial u_r}{\partial r} + \frac{1}{\rho}\frac{\partial p}{\partial r} - \nu \left( \frac{ \partial^2 u_r }{\partial y^2} + \frac{1}{r}\frac{\partial}{\partial r} ( r\frac{\partial u_r}{\partial r} ) + \frac{1}{r^2}\frac{\partial^2 u_r}{\partial \theta^2} - \frac{u_r}{r^2} \right) = \Omega^2 r.
\label{eq:momentum_polars_simple}
\end{equation}
Note that, in this numerical simulation, we have $\Omega=1$, $\rho=1$ and $\nu = 1/125$, in the normalized units.
Let us now estimate the order of magnitude of the remaining terms.
Sufficiently far from the wing tip, i.e., for $r<0.65R$, the spanwise velocity varies approximately like $u_r = A \Omega r$, where $A$ is of order unity, assumed constant for a rough estimate.
Then we obtain $u_r \frac{\partial u_r}{\partial r} \approx \Omega^2 r$. At the same location, the pressure decreases with $r$.
From figure~\ref{fig:pressure}(\textit{b}), the pressure gradient term can be estimated as $\nabla p / \rho = - B \Omega^2 r$, where $B$ is of order unity.
Finally, figure~\ref{fig:velocity}(\textit{c}) indicates that, at $r=R/2=2$, the vertical size of the bubble is of order $0.1$ of the local chord length, suggesting that the second derivative of the velocity is of order 100 in the bubble.
Computation shows that $\partial^2 u_r / \partial y^2 \approx -300$. Therefore, the viscous term is also of order unity.
Hence, we conclude that the spanwise flow and the negative radial pressure gradient are due to the centrifugal force and viscous stresses.

\subsection{Parametric study of vortex shedding regimes}\label{Sec:parametric}

It is of practical interest to determine the marginal value of the
aspect ratio $\Lambda_c$ such that no vortex shedding occurs if
$\Lambda<\Lambda_c$. A series of numerical simulations have been
carried out to estimate this value. In all of these computations,
the domain size was set to $L^3=16.8^3$ and the number of grid points
to $N^3=512^3$. In the range of $\alpha$ between 30 and 60 degrees, vortex shedding occurs at $\Lambda>\Lambda_c = 6$,
and this value is almost independent of $\alpha$.
When the angle of attack is
sufficiently small, there is no vortex shedding regardless of the
aspect ratio.

\begin{figure}
\centering
\includegraphics{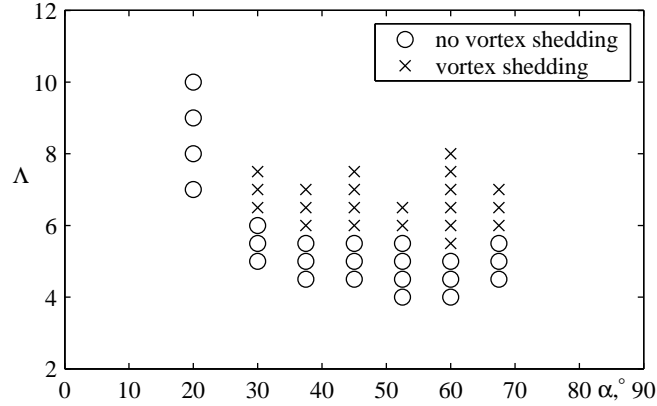}
\caption{Flow regimes observed in the simulations, depending on
the angle of attack and aspect ratio.}\label{fig:chart_sum}
\end{figure}

The results are summarized in
figure~\ref{fig:chart_sum}, which depicts the flow regime observed
in each simulation.
The absolute value of the time derivative of the lift force coefficient, $|\mathrm{d}c_L/\mathrm{d}t|$,
serves to indicate vortex shedding: if it exceeds 0.5 during the second half-revolution ($\pi < \theta < 2\pi$),
the flow regime is named ``shedding'', otherwise it is called ``no shedding''.
This is a crude classification.
Note that the flow is three-dimensional,
therefore the wing tip flow separation also has some effect on the
time evolution of the aerodynamic force.
This effect is small when $\alpha<60^\circ$ (for example, see the discussion of the flow and the forces at $\alpha=30^\circ$ in the previous section).
However, at larger $\alpha$
it becomes comparable with the effect of vortex shedding from the leading and
trailing edges. Therefore, at $\alpha>60^\circ$, the criterion based on
$\mathrm{d}c_L/\mathrm{d}t$ fails, hence these regimes are not shown in figure~\ref{fig:chart_sum}.
Another possible source of unsteadiness is the spiral tip vortex. In the present study,
we minimize its effect by stopping the simulations after the wing makes one complete revolution and encounters its own wake.


\subsection{Inviscid fluid model of spanwise flow}\label{Sec:inviscid}

In section~\ref{Sec:dynamics_forces}, we discussed the spanwise flow in the recirculation bubble and pointed out the significance of viscous stresses in it.
Along with that discussion, it is instructive to consider a simple potential flow model of a similar flow,
as it provides an analytical closed form expression of the velocity field.
It describes, qualitatively, the flow outside of the recirculation bubble and vortex sheets, where the flow is irrotational.
For instance, it predicts a reverse spanwise flow (from wing tip to root) in the neighbourhood of the front stagnation point.
This effect is seen in the numerical simulation in figure~\ref{fig:velocity}(\textit{c}).

\begin{figure}
\begin{center}
\includegraphics[height=4.2cm,clip]{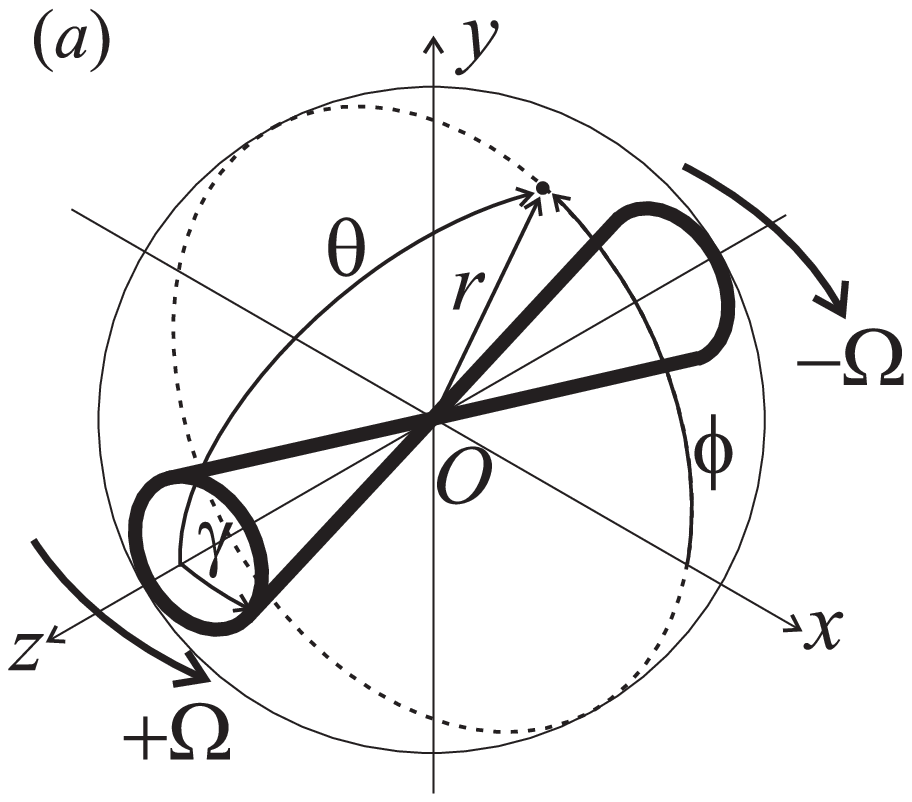}
\quad\quad\quad
\includegraphics[height=4.2cm,clip]{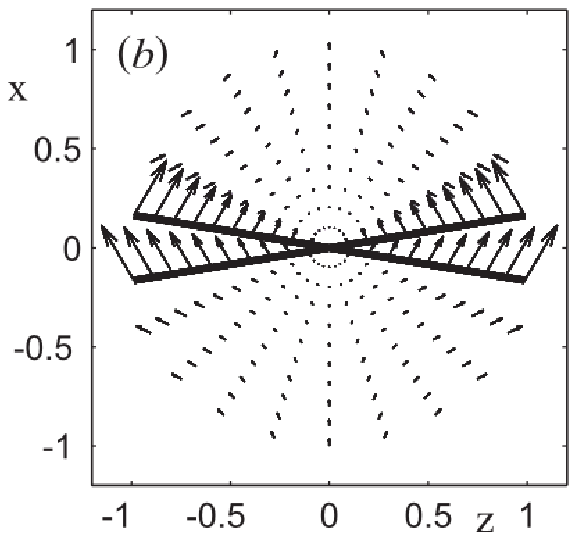}
\end{center}
\caption{\label{fig_cones} Schematic of two counter-rotating cones
(\textit{a}) and the potential velocity in the horizontal plane
(\textit{b}).}
\end{figure}

Let us consider the flow about two infinite coaxial cones that
rotate in opposite directions, as shown in
figure~\ref{fig_cones}(\textit{a}). Here $\Omega$ is the angular
velocity, $\gamma$ is the angle at the vertex of the cone, and
$r,\theta,\phi$ are the polar coordinates.
The geometry parameter $\gamma$ plays a role similar to that of $\Lambda$
in the previous section: it shows how fast the chord length (in this case the diameter) increases with the radius $r$.

The velocity potential $\Phi(r,\theta,\phi)$ satisfies the
following boundary value problem for Laplace's equation:
\begin{equation}\label{eq:laplace}
\begin{array}{ll}
\displaystyle \nabla^2 \Phi = 0, & \quad r \in ]0,\infty[,~ \theta \in ]\gamma,\pi-\gamma[, ~ \phi \in [0,2\pi[, \\
\displaystyle \frac{\partial \Phi}{r \partial \theta} = \Omega r \cos \phi, & \quad \textrm{on}~ \theta=\gamma, \\
\displaystyle \frac{\partial \Phi}{r \partial \theta} = - \Omega r \cos \phi, & \quad \textrm{on}~ \theta=\pi-\gamma, \\
\Phi = 0 & \quad \textrm{at}~ r=0.
\end{array}
\end{equation}
The solution to (\ref{eq:laplace}) can be found in terms of the Legendre function $Q_2^1$,
\begin{equation}\label{potential}
\Phi(r,\theta,\phi) = \frac{\Omega r^2
Q_2^1(\cos\theta)\cos\phi}{\mathrm{d}Q_2^1(\cos\gamma)/\mathrm{d}\gamma},
\end{equation}
where
\begin{equation}\label{eq:legendre_function}
Q_2^1(\cos\theta) =
-\frac{3}{2}\cos\theta\sin\theta\ln\frac{1+\cos\theta}{1-\cos\theta}+2\sin\theta-\frac{\cos^2\theta}{\sin\theta},
\end{equation}
Figure~\ref{fig_cones}(\textit{b}) displays the velocity $\bm{u}=\nabla\Phi$ in the
horizontal plane. Its radial component is outwards behind the
cones and inwards in front of them, in agreement with the
numerical simulation. The radial component, normalized to $\Omega r$, is
\begin{equation}\label{eq:radial_velocity_inviscid}
\frac{u_r}{\Omega r}
=
\frac{2Q_2^1(\cos\theta)\cos\phi}{\mathrm{d}Q_2^1(\cos\gamma)/\mathrm{d}\gamma}.
\end{equation}
Its maximum value depends on the cone angle $\gamma$
at the vertex of the cone 
almost linearly when $\gamma<\pi/10$,
\begin{equation}\label{radial_velocity}
u_r \approx -2 \gamma \Omega r \cos\phi
~~~\mathrm{on}~~\theta=\gamma.
\end{equation}

At the front stagnation point, $\phi=0$,
the spanwise flow is towards the centre of rotation.
For $\gamma=\pi/12$, which is about the same angle as for our smallest-aspect-ratio wing $\Lambda=4$ discussed in section~\ref{Sec:dynamics_forces},
the velocity at $\phi=0$, $r=2$, $\Omega=1$ is equal to $u_r \approx 1$, which is somewhat larger, but comparable with, the maximum inward flow velocity displayed in figure~\ref{fig:velocity}(\textit{c}).
Of course, in a viscous flow, the velocity at the boundary is zero due to the no-slip condition, and the maximum velocity reached at the edge of the boundary layer
is therefore smaller than that given by the potential flow theory.
Hence, we conclude that the inward spanwise flow near the front stagnation point is a potential flow effect,
unlike the outward flow in the recirculation bubble, which is a combined effect of inertia, incompressibility and viscosity.


\subsection{Aspect ratio effect at a constant span-based Reynolds number}\label{Sec:reynolds}

The numerical simulations presented in the previous sections have been carried out at a constant mid-chord-based Reynolds number $Re$.
However, when the spanwise velocity component is large, some features of the flow field may be controlled by the span-based Reynolds number $Re_R=\Omega R^2 / \nu = 2 \Lambda Re$ \cite{Harbig_etal_2013}.
In order to clarify the effect of $Re_R$ on the transition to vortex street shedding,
this section presents some results of a numerical simulation of the flow past a low aspect ratio wing ($\Lambda=4$) at $Re=1000$, which
yields $Re_R=8000$. Note that the same value of $Re_R$ corresponds to the flow past a high aspect ratio wing ($\Lambda=16$) discussed in section~\ref{Sec:dynamics_forces}.
The numerical simulation was carried out using a finer discretization grid of $768^3$ points, as required for resolving small vortical structures due to the increased Reynolds number.
The angle of attack was fixed to $\alpha=30^\circ$, as in section~\ref{Sec:dynamics_forces}.

\begin{figure}
\centering
\includegraphics[scale=0.72]{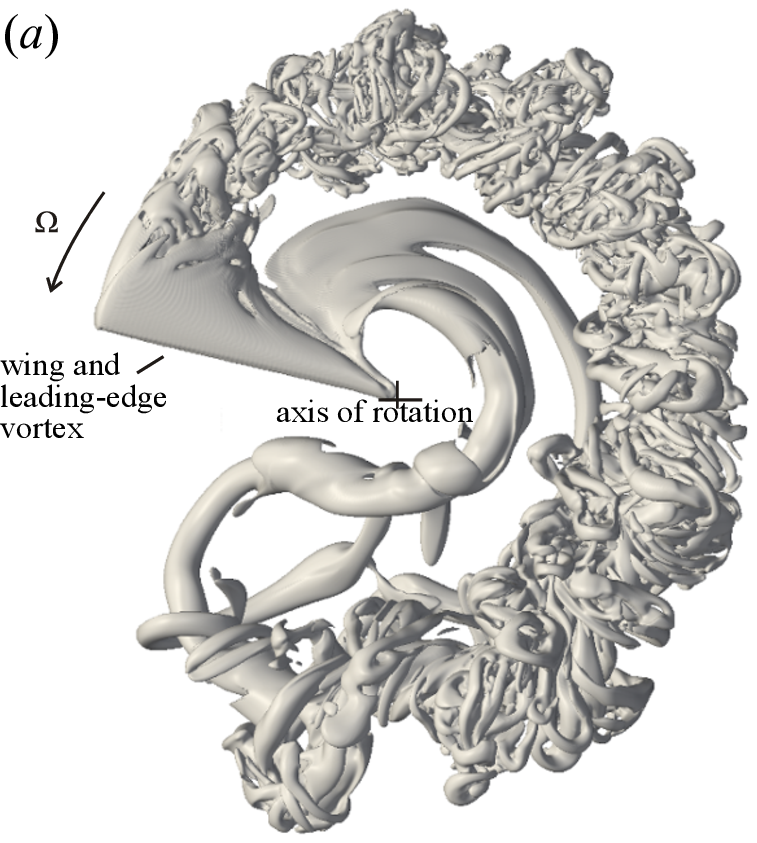}
\includegraphics[scale=0.67]{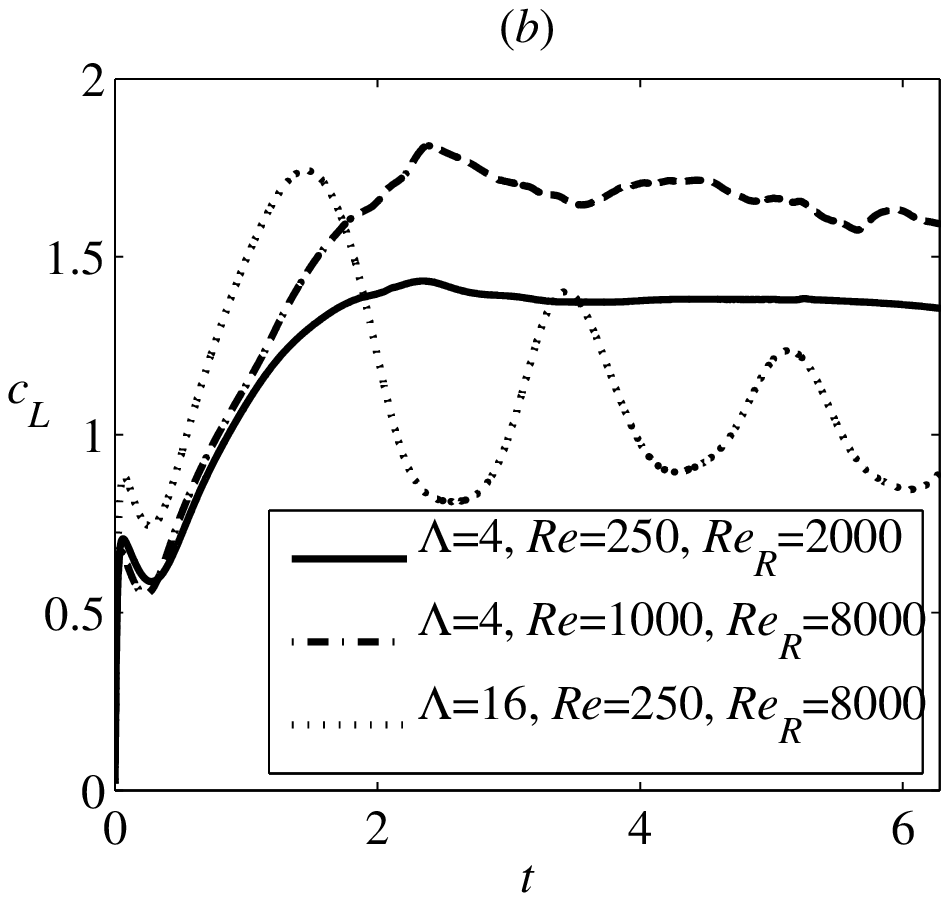}
\caption{(\textit{a}) Isosurfaces of the $\lambda_2$-criterion for $\Lambda=4$, $Re=1000$ at time $t=6.28$. (\textit{b}) Influence of $Re$ and $Re_R$ on the time evolution of the lift coefficient.} \label{fig:re1000}
\end{figure}

Figure~\ref{fig:re1000}(\textit{a}) displays an iso-surface of $\lambda_2=-0.01$.
The presence of small-scale structures in the tip vortex and partly in the leading edge vortex
is the main difference from the lower Reynolds number case $Re=250$, $\Lambda=4$, discussed in section~\ref{Sec:dynamics_forces}.
However, there are no signs of alternate vortex shedding from the leading/trailing edges and there is no vortex street in the wake, in contrast to the case $Re=250$, $\Lambda=16$.

Time evolution of $c_L$ is shown in figure~\ref{fig:re1000}(\textit{b}).
The new case $Re=1000$, $\Lambda=4$ is shown with a dashed line. Let us first compare it to the same wing operating at $Re=250$.
For $t<1$, there is little sensitivity of $c_L$ to the change in $Re$. However, the values of $c_L$ attained after
the initial transient are about 20\% larger in the higher Reynolds number case.
The extra lift can be explained by the pressure deficit generated in the leading edge vortex core, which is more compact at larger $Re$ \cite{Shyy_Liu_2007,Maxworthy_2007}.
This effect is also consistent with earlier experiments \cite{Birch_etal_2004,Lentink_Dickinson_09b}.

There are some noticeable fluctuations in the time evolution of $c_L$ at $Re=1000$.
However, their amplitude is small and there is no dominant frequency, in contrast to the case $\Lambda=16$, $Re=250$.
Therefore, transition to the vortex street type of wake is controlled by the aspect ratio $\Lambda$ and the chord-based Reynolds number $Re$
in a way that cannot be reduced to just considering the span-based Reynolds number $Re_R=2 \Lambda Re$.

\section{Conclusion and perspectives}\label{Sec:conclusion}

Numerical simulation of flows past revolving wings have been carried out for a series of values of aspect ratio $\Lambda$ and angle of attack $\alpha$, with the Reynolds number fixed at $Re=250$.

It was found that the structure of the vortex wake significantly depends on $\Lambda$.
If $\Lambda>6$ and $\alpha>20^\circ$, radial vortices are shed from most part of the
leading edge as well as most part of the trailing edge. This vortex shedding is accompanied by high amplitude oscillation of the aerodynamic force acting on the wing. If $\Lambda<5.5$, no vortex shedding occurs and the wing generates a
stable leading-edge vortex found in earlier studies \citeaffixed{Maxworthy_79,Liu_Kawachi_98}{\eg}.

In a steady recirculation bubble past the wing,
spanwise velocity in the outward direction and negative pressure gradient appear due to the action of centrifugal force and viscous stresses.
This velocity is of the same order of magnitude as the wing circumferential velocity.
Near the front stagnation point, the spanwise velocity is in the opposite direction, from wing tip to root,
and this is purely a potential flow effect due to incompressibility.

The structure of the leading-edge vortex also depends on the Reynolds number, therefore we only expect the above conclusions to hold if $Re \approx 250$. Exploring the dependence of $\Lambda_c$ on the Reynolds number is a possible topic for future work.

\ack

Numerical simulations were carried out using HPC resources of IDRIS,
Paris, project 81664.


\section*{References}




\end{document}